\documentclass[twocolumn, preprintnumbers, showpacs, nofootinbib, aps, prl]{revtex4-1}
\usepackage{graphicx}

\newcommand{\bi}{\bibitem}
\newcommand{\be}{\begin{eqnarray}}
\newcommand{\ee}{\end{eqnarray}}

\begin{document}

\title{Spinning super-massive objects in galactic nuclei up to $a_* > 1$}

\author{Cosimo Bambi}
\email{cosimo.bambi@ipmu.jp}

\affiliation{
Institute for the Physics and Mathematics of the Universe, 
The University of Tokyo, Kashiwa, Chiba 277-8583, Japan}

\date{\today}

\preprint{IPMU11-0002}

\begin{abstract}
Nowadays we believe that a typical galaxy contains about $10^7$ 
stellar-mass black holes and a single super-massive black hole at 
its center. According to general relativity, these objects are 
characterized solely by their mass $M$ and by their spin parameter 
$a_*$. A fundamental limit for a black hole in general relativity
is the Kerr bound $|a_*| \le 1$, but the accretion process can 
spin it up to $a_* \approx 0.998$. If a compact object is not a
black hole, the Kerr bound does not hold and in this letter 
I provide some evidences suggesting that the accretion 
process could spin the body up to $a_* > 1$. 
While this fact should be negligible for stellar-mass objects, 
some of the super-massive objects at the center of galaxies may 
actually be super-spinning bodies exceeding the Kerr bound. Such 
a possibility can be tested by gravitational wave detectors
like LISA or by sub-millimeter very long baseline interferometry
facilities.
\end{abstract}

\pacs{04.50.Kd, 04.70.Bw, 97.60.Lf}

\maketitle


{\it Introduction ---}
There are robust observational evidences of the existence of $5-20$
Solar mass compact bodies in X-ray binary systems~\cite{rev1} 
and of $10^5-10^9$ Solar mass bodies at the center of most 
galaxies~\cite{rev2}. All these objects are commonly interpreted 
as black holes (BHs) because they cannot be explained otherwise 
without introducing new physics. In particular, stellar-mass objects
in X-ray binary systems are too heavy to be neutron or quark stars 
for any reasonable matter equation of state~\citep{ruf}. 
Observations of stellar orbits around the super-massive BH candidate 
at the center of the Galaxy show that this object is too massive, 
compact, and old to be a cluster of non-luminous bodies~\citep{maoz}.

In 4-dimensional general relativity, BHs are known as Kerr BHs
and are completely specified by their mass, $M$, and by their 
dimensionless spin parameter, $a_* = J/M^2$, where $J$ is the 
BH spin angular momentum. The mass $M$ sets the size of the 
system and, in principle, it can assume any value. The spin 
parameter $a_*$ determines the properties of the space-time. 
It must satisfy the constraint $|a_*| \le 1$, which is the 
condition for the existence of the event horizon.

In general, the value of the spin parameter is determined by the
competition of three physical processes: the event creating the 
object, mergers, and gas accretion. Recent simulations taking 
the relevant microphysics processes into account find that the 
BH formed after the collapse of a super-massive star has 
$a_* \sim 0.5 - 0.8$, and then it is spun up to $a_* \sim 0.6 - 
0.9$, depending on the initial angular velocity of the collapsing 
stellar core~\cite{sh10}. In the case of the merger of two 
neutron stars, the product of the coalescence is a BH with 
$a_* \approx 0.78$, depending only weakly on the total mass 
and mass ratio of the system~\cite{sh09}. The capture of small 
bodies (minor mergers) in randomly oriented orbits typically 
spins a BH down~\cite{hughes,gammie}, because the magnitude of 
the orbital angular momentum for corotating orbits is smaller 
than the one for couterrotating orbits. The case of coalescence 
of two BHs with comparable mass (major merger) has been addressed 
only in the last few years~\cite{pret}. For random 
mergers, the most probable final product is a BH with $a_* 
\approx 0.70$, while fast-rotating objects with $a_* > 0.9$ 
should be rare~\cite{berti}.

The gas in an accretion disk falls to 
the BH by loosing energy and angular momentum. When it 
reaches the innermost stable circular orbit (ISCO), it is 
quickly swallowed by the BH, which changes its mass by
$\delta M = \epsilon_{\rm ISCO} \delta m$ and its spin by
$\delta J = \lambda_{\rm ISCO} \delta m$, where $\epsilon_{\rm ISCO}$ 
and $\lambda_{\rm ISCO}$ are respectively the specific energy 
and the specific angular momentum of a test-particle at the
ISCO, while $\delta m$ is the gas rest-mass. The evolution 
of the spin parameter is thus governed by the following 
equation~\cite{bardeen}
\be\label{eq-a}
\frac{da_*}{d\ln M} = \frac{1}{M} 
\frac{\lambda_{\rm ISCO}}{\epsilon_{\rm ISCO}} - 2 a_* \, .
\ee
If the disk is on the BH equatorial plane, the equilibrium is
reached for $a_*^{eq} = 1$ by swallowing a finite amount of 
matter. For example, an initially non-rotating BH reaches the 
equilibrium after increasing its mass by a factor $\sqrt{6} 
\approx 2.4$~\cite{bardeen}. Including the effect of the
radiation emitted by the disk and captured by the BH, one
finds $a_*^{eq} \approx 0.998$~\cite{thorne}, because radiation
with angular momentum opposite to the BH spin has larger 
capture cross section. The presence of magnetic fields in 
the plunging region may further reduce this value to $a_*^{eq} 
\sim 0.95$~\cite{gammie,sh05}, by transporting angular momentum 
outward.

For stellar-mass BHs, the processes of mergers and gas accretion 
are more likely negligible. If they belong to low-mass X-ray 
binary systems, even swallowing the stellar companion they cannot
change significantly their spin. If they are in high-mass 
X-ray binary systems, even accreting at the Eddington limit they
do not have enough time to grow before the explosion of the 
companion. So, the value of the spin of stellar-mass BHs should 
reflect the one at the time of their creation. On the contrary, 
for super-massive objects the spin is determined by their evolution
history, since they have increased their mass by a few orders of 
magnitude from the original value. For prolonged disk accretion, 
the BH has the time to align itself with the disk, the process 
of gas accretion dominates over mergers, and the low-redshift 
super-massive BH population in galactic nuclei would be 
characterized by fast-rotating objects, $a_* > 0.9$, with some of 
them possibly close to the theoretical bound 
$a_* \approx 0.998$~\cite{marta,berti}.

Since there are no clear evidences that the current BH candidates 
are really the BH predicted by general relativity, it is worth
investigating other possibilities. 
In this letter I discuss the accretion process from a thin disk
in a space-time with deviations from the Kerr metric and I argue
that the central object may be spun up to $a_* > 1$.
Such a possibility is a new finding, but it does not contradict 
any rule or belief, because the Kerr bound $|a_*| \le 1$ is defined 
only for BHs. 
This possibility is well known for objects such as main-sequence
stars and white dwarfs. However, it does not seem so trivial that this 
is also the case for extremely compact objects.
The key ingredient is that any deviation from the 
Kerr geometry makes the inner radius of the disk of very 
fast-rotating objects significantly larger than the one of BHs.
In most cases, this leads to a higher value of the quantity
$\lambda_{\rm ISCO}/\epsilon_{\rm ISCO}$ in Eq.~(\ref{eq-a}) and 
may move 
$a_*^{eq}$ beyond 1. The super-massive objects in galactic 
nuclei may thus be super-spinning objects violating the Kerr 
bound~\cite{io}.

{\it Non-Kerr compact objects ---}
The Manko-Novikov (MN) metric is a stationary, axisymmetric, and 
asymptotically flat exact solution of the vacuum Einstein 
equations~\cite{mn}. It is not a BH solution, but it can be used 
to describe the gravitational field outside a generic body. The 
expression of the metric is quite long and can be seen 
in~\cite{baba}, where we corrected a few typos of the original 
article by Manko and Novikov. The solution has an infinite number 
of free parameters, which determine the mass, the spin, and all 
the higher order mass multipole moments of the gravitational field.
For the sake of simplicity, here I consider only three parameters,
determining the mass $M$, the spin $J$, and the quadrupole moment 
$Q$. The latter can be parametrized by the anomalous quadrupole 
moment $q$, defined by
\be
Q = Q_{\rm Kerr} - q M^3 \, ,
\ee
where $Q_{\rm Kerr} = - a_*^2M^3$ is the quadrupole moment of a 
BH. For $q=0$ we recover exactly the Kerr metric, while for $q>0$
($q<0$) the object is more oblate (prolate) than a BH. Another
interesting parametrization is
\be\label{eq-qt}
Q = - (1 + \tilde{q}) a_*^2 M^3 \, ,
\ee
presumably with $\tilde{q} \ge -1$, since it is difficult to 
imagine that the rotation makes the object more and more prolate.

\begin{figure}
\par
\begin{center}
\includegraphics[height=6cm,angle=0]{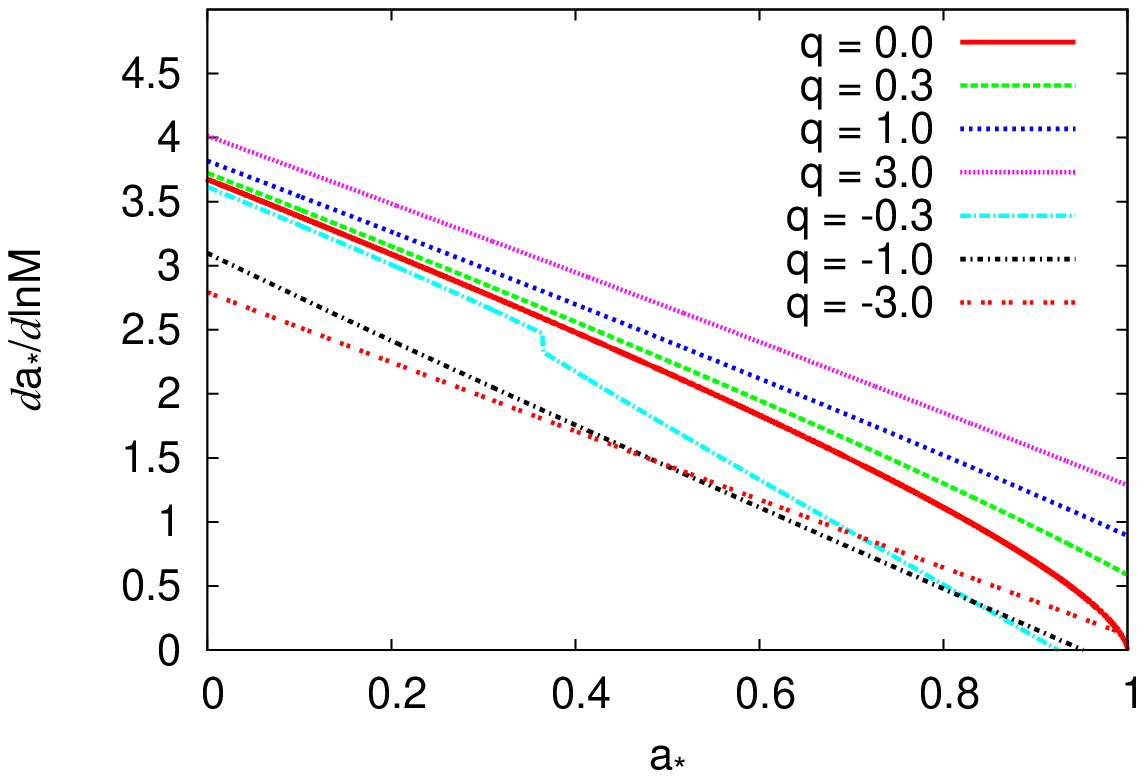}
\includegraphics[height=6cm,angle=0]{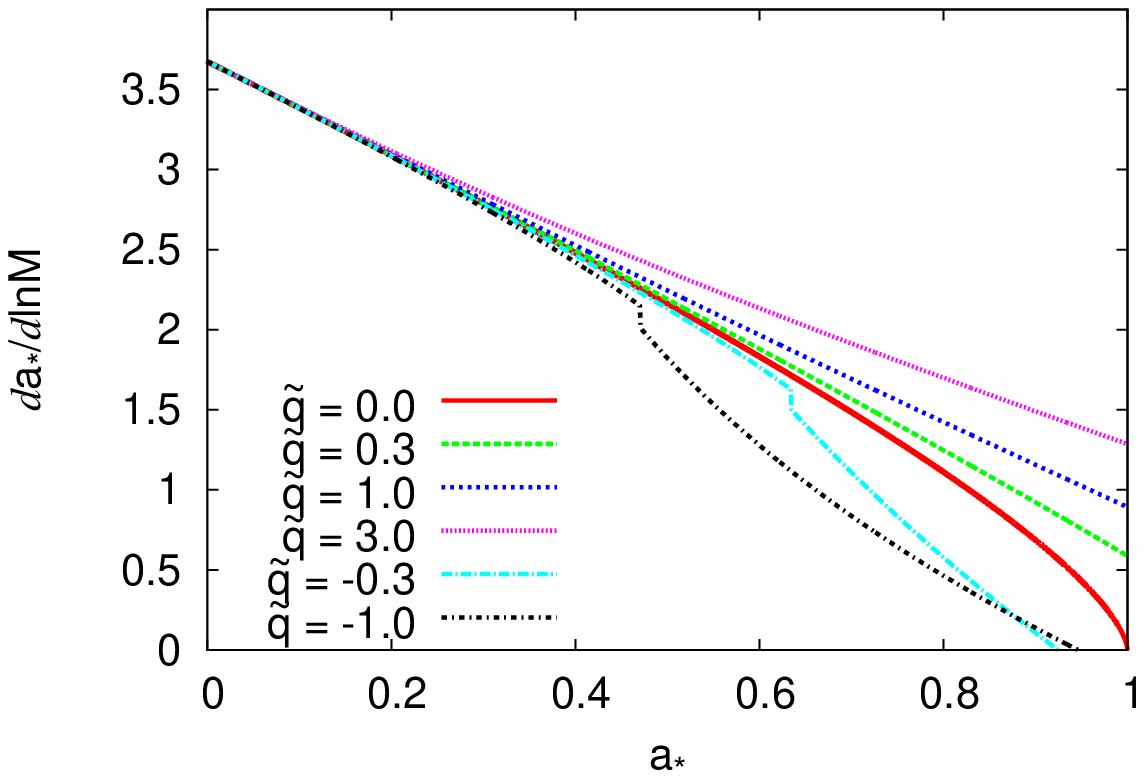}
\end{center}
\par
\vspace{-5mm} 
\caption{$d a_*/d \ln M$ as a function of the spin parameter $a_*$ 
for several values of the anomalous quadrupole moment $q$ (upper 
panel) and $\tilde{q}$ (lower panel). As long as $d a_*/d \ln M > 0$, 
the gas of the disk spins the object up. The fact that 
$d a_*/d \ln M > 0$ at $a_* = 1$ suggests that the equilibrium is 
reached for $a_*^{eq} > 1$}, which is impossible for black holes.
\label{fig1}
\end{figure}

For a disk on the equatorial plane, $\epsilon_{\rm ISCO}$
and $\lambda_{\rm ISCO}$ can be computed numerically as
described in~\cite{baba} and then one can evaluate $da_*/d\ln M$.
The result is plotted in Fig.~\ref{fig1}. For objects more
oblate than a BH, the right hand side of Eq.~(\ref{eq-a})
is always positive, suggesting that 
$a^{eq}_* > 1$. For objects
more prolate than a BH, one finds two different cases.
When $q \gtrsim -2$, $da_*/d\ln M$ becomes null before $a_* = 1$: 
in this case $a_*^{eq} < 1$, with a minimum $a_*^{eq} \approx 0.92$ 
at $q \approx -0.3$. When $q \lesssim -2$, $da_*/d\ln M$ is 
always positive and presumably $a_*^{eq} > 1$.

\begin{figure}
\par
\begin{center}
\includegraphics[height=6cm,angle=0]{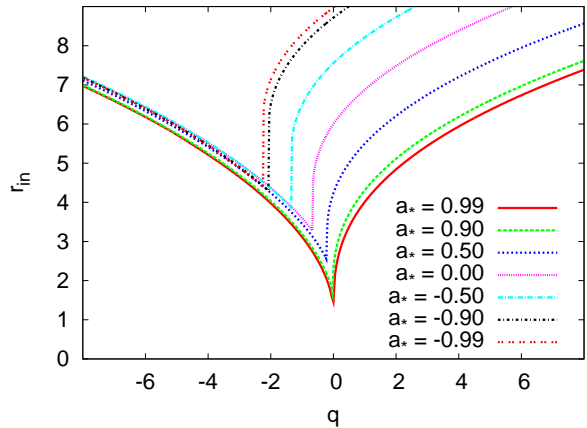}
\end{center}
\par
\vspace{-5mm} 
\caption{Inner radius of the disk in Schwarzschild coordinates as 
a function of the anomalous quadrupole moment $q$ (the case $q=0$
corresponds to a black hole) for different values of the
spin parameter $a_*$. Inner radius in units $M=1$.}
\label{fig2}
\end{figure}

Fig.~\ref{fig1} can be interpreted by considering the behavior of
the inner radius of the disk, see Fig.~\ref{fig2} and, for
more details, Appendix~B of Ref.~\cite{baba}. It turns out that
the inner radius of the disk of generic fast-rotating objects is
always larger than in the BH case and this is roughly reflected
in a higher value of $\lambda_{\rm ISCO}/\epsilon_{\rm ISCO}$.
For $q > 0$, the inner radius can be found as in the Kerr metric.
For $q < 0$, the picture is more complex. For any spin parameter 
$a_*$, there are two critical values, say $q_1$ and $q_2$ with 
$q_1>q_2$. If $q>q_1$, there are no differences with the $q\ge0$ case.
If $q_2<q<q_1$, there are two disconnected regions with stable
circular orbits: the standard region $r>r_1$ and an internal
one with $r_3<r<r_2$ ($r_2<r_1$). Since the orbits in the
internal region have higher energy and angular momentum than the
orbit at $r_1$, the gas cannot go from $r_1$ to an orbit in the
internal region, and the inner radius of the disk is thus $r_1$.
As $q$ decreases, $r_1$ and $r_2$ approaches each other and, 
for $q \le q_2$, we have only one region with internal radius
$r_3$. At $q=q_2$ we have thus a sudden decreases of the inner
radius of the disk, as we can see in Fig.~\ref{fig2}. The
discontinuities of the curves in Fig.~\ref{fig1} for $q = -0.3$ 
and $\tilde{q} = -0.3$ and $-1.0$ correspond to this transition. 
For $q = -1.0$ and $-3.0$, the transition occurs at $a_*<0$. Let 
us notice that when $a_*$ approaches 1, $q_1$ and $q_2$ approach 0.

Here I have only shown the case of objects whose gravitational 
field has quadrupole deformations with respect to the Kerr
metric. However, the result seems to be much more general. It turns
out that only in an exact Kerr metric the radius of the ISCO 
goes to $M$ and that any deviation from the Kerr background 
makes the inner radius of the disk significantly larger. While 
this does not strictly mean that $a_*^{eq}>1$ for non-Kerr 
objects (see the case $-2\lesssim q < 0$), the space region 
with $a_*^{eq} < 1$ is always limited. 
The reason of this peculiarity of the Kerr metric is
presumably related to the fact that only in this special 
case the space-time is ``regular''. For $q \neq 0$, there
are some pathological features (e.g. closed time-like curves) 
at very small radii: they can be neglected in our study, 
because they are inside the inner radius, and the idea is 
indeed that this region does not exist in 
reality, because inside the object, while the MN
solution would describe the exterior gravitational field.
Since the inner radius of the disk turns out to be always outside 
the pathological region, except in the case of an exact Kerr
metric it can never goes to $M$ as $a_*$ approaches 1.

{\it Discussion ---}
When the right hand side of Eq.~(\ref{eq-a}) is positive,
the accreting gas spins the object up. 
If we admit deviations from the Kerr geometry, 
$da_*/d\ln M$ can be positive at $a_* = 1$ (while it is zero for 
a BH) and this suggests the possibility of an equilibrium
spin parameter exceeding the Kerr bound.
Unfortunately, the
MN solution requires $|a_*| < 1$ and therefore it is
impossible to predict $a_*^{eq}$. 
However, this is likely only a problem of coordinates.
The same problem exists for the Kerr case: the 
MN solution is in prolate quasi-cylindrical coordinates and 
requires $|a_*| < 1$, while if we use Boyer-Lindquist
or Kerr-Schild coordinates we can consider objects with $|a_*| >1$,
even if they are not BHs but naked singularities.

Since we do not know the nature of this objects, we cannot 
say which value of $q$ is reasonable. We can notice 
that for any common equation of state a rotating object is 
more oblate than a BH (that is, $q$ and $\tilde{q} > 0$) 
and that for a neutron star one expect a quadrupole moment $Q$ 
as in Eq.~(\ref{eq-qt}), with $\tilde{q} \approx 1 - 10$, 
depending on the equation of state 
and the mass of the object~\cite{neutron}. So, it is not
unreasonable to expect sizable deviations from the theoretical
bound for BHs $a_* \approx 0.998$, even including the (small) 
effect of capture of radiation emitted by the disk~\cite{thorne} 
and the (more important) one of magnetic fields in the 
plunging region~\cite{gammie,sh05}.

In the case of non-Kerr objects, we can neither predict the natal 
spin, nor the outcome of major mergers, because we do not know
their internal structure. Like for BHs, minor mergers
spin the object down, since the magnitude of the angular
momentum of particles in counterrotating orbits is larger than
the one of particles in corotating orbits. However, for 
prolonged disk accretion, the other mechanisms are not important 
as long as the timescale of the alignment of the spin of the 
object with the disk is much shorter than the time for the mass 
to increase significantly. For a BH, the alignment timescale 
is determined by the coupling between its spin and the gas
orbital angular momentum and is~\cite{bp1}
\be\label{eq-al}
t_{\rm align} \sim \frac{a_* M^{3/2}}{\dot{M} R_W^{1/2}}
\left(\frac{\nu_1}{\nu_2}\right) \, ,
\ee
where $\dot{M}$ is the mass accretion rate, $R_W \sim 10^4 \, M$
is the warp radius, while $\nu_1$ and $\nu_2$ are respectively the
viscosities acting in the plane of the disk and normal to the
disk. For a thin disk, $t_{\rm align}$ can be much shorter than the 
accretion timescale $t_{\rm acc} \sim M/\dot{M}$~\cite{bp2,berti}.
The same estimate holds for a generic body as long as 
$|\tilde{q}| \ll 100$, because $t_{\rm align}$ is still determined by the
spin: the angular frequency of the precession induced by the 
spin is $\omega_J \sim J r^{-3}$, while the one induced by
the quadrupole moment is $\omega_Q \sim Q M^{-1/2} r^{-7/2}$.

Till now, I neglected the possibility that these objects
become unstable before reaching $a_*^{eq}$. At least for small 
deviations from the Kerr metric, one can expect instabilities
for $a_*$ a bit larger than 1 if the object is so compact to have 
an external ergoregion~\cite{barausse}. The object may also 
become unstable before the spin parameter reaches 1, depending 
on its equation of state, just like a neutron star can never
spin up to a rotational period shorter than about 1~ms due to
r-mode instabilities~\cite{kokkotas}. Since the process of 
spin-up by the accreting gas is unavoidable in most cases, 
there are two possible answers: $i)$ these objects are stable
(either because deviations from the Kerr metric are not small 
or because they are not surrounded by an ergoregion) or $ii)$ 
they are indeed unstable. In the latter case, they should 
decay (probably spin down, as neutron stars do~\cite{kokkotas}) 
to a stable configuration emitting gravitational waves. If 
we consider the large number of expected super-massive objects 
in our Universe, there might exist a relevant background of 
low-frequency ($\nu \lesssim 1/M$) cosmic gravitational waves.

The possibility that most of the super-massive objects in
galactic nuclei are not BHs, but super-spinning bodies, is quite
relevant and cannot be ignored in view of future experiments
like the gravitational wave detector LISA or sub-millimeter
very long baseline interferometry facilities. In the literature,
there are several works discussing how we can test the Kerr
nature of these super-massive objects, see e.g. Refs.~\cite{bumpy,
vsop}. For example, by observing the inspiral of a 10~Solar
masses object into a 10$^6$~Solar masses object for one year, LISA
will be able to constrain the quadrupole moment of the
super-massive body with a precision of $10^{-3}$~\cite{barack}.
Surprisingly, all these works always assume that $|a_*|<1$.
On the other hand, the conclusion of this letter is that if we 
consider the possibility that these objects are not BHs, we 
cannot restrict our attention to the case $|a_*|<1$.

{\it Conclusions ---}
BHs in general relativity must satisfy the bound $|a_*| \le 1$,
where $a_*$ is the spin parameter. This is just the condition for
the existence of the event horizon. The accretion process can spin 
a BH up to $a_* \approx 0.998$ and we currently believe that most 
of the super-massive compact objects in galactic nuclei are
rapidly rotating super-massive BHs. In this letter I suggested
that deviations from the Kerr metric may have the accretion process 
spin the compact object up to $a_* > 1$. If the super-massive
objects at the center of galaxies are not the black holes 
predicted by general relativity, their spin parameter could
violate the constraint $|a_*| \le 1$. Unfortunately, we
cannot predict the equilibrium spin parameter for an object 
with a given quadrupole moment, because the MN solution is in
quasi-cylindrical coordinates and requires $|a_*|<1$. An
extension to include super-spinning objects is definitively 
non-trivial, but might not be impossible~\cite{mm},
at least numerically.

The good news of this story is that an anomalous quadrupole 
moment may induce an anomalously high value of the spin, whose
general relativistic effects would be larger and thus 
easier to discover.

\begin{acknowledgments}
I would like to thank Enrico Barausse and Naoki Yoshida
for critically reading a preliminary version of this manuscript 
and providing useful feedback. This work was supported by World 
Premier International Research Center Initiative (WPI Initiative), 
MEXT, Japan, and by the JSPS Grant-in-Aid for Young Scientists 
(B) No. 22740147.
\end{acknowledgments}




\begin{thebibliography}{99}

\bibitem{rev1}
  A.~P.~Cowley,
  Ann.\ Rev.\ Astron.\ Astrophys.\  {\bf 30}, 287 (1992).

\bibitem{rev2}
  J.~Kormendy and D.~Richstone,
  Ann.\ Rev.\ Astron.\ Astrophys.\  {\bf 33}, 581 (1995).

\bibitem{ruf}
  C.~E.~.~Rhoades and R.~Ruffini,
  Phys.\ Rev.\ Lett.\  {\bf 32}, 324 (1974);
  V.~Kalogera and G.~Baym,
  Astrophys.\ J.\  {\bf 470}, L61 (1996).

\bibitem{maoz}
  E.~Maoz,
  Astrophys.\ J.\  {\bf 494}, L181 (1998).

\bibitem{sh10}
  Y.~Sekiguchi and M.~Shibata,
  arXiv:1009.5303 [astro-ph.HE].

\bibitem{sh09}
  K.~Kiuchi, Y.~Sekiguchi, M.~Shibata and K.~Taniguchi,
  Phys.\ Rev.\  D {\bf 80}, 064037 (2009).

\bibitem{hughes}
  S.~A.~Hughes and R.~D.~Blandford,
  Astrophys.\ J.\  {\bf 585}, L101 (2003).

\bibitem{gammie}
  C.~F.~Gammie, S.~L.~Shapiro and J.~C.~McKinney,
  Astrophys.\ J.\  {\bf 602}, 312 (2004).

\bibitem{pret}
  F.~Pretorius,
  in {\it Physics of Relativistic Objects in Compact Binaries: from Birth to Coalescence}, 
  edited by Colpi et al.
  (Springer Verlag and Canopus Publishing Limited, Bristol, UK, 2009),
  pp. 305-369.

\bibitem{berti}
  E.~Berti and M.~Volonteri,
  Astrophys.\ J.\  {\bf 684}, 822 (2008).

\bibitem{bardeen}
  J.~M.~Bardeen,
  Nature {\bf 226}, 64 (1970).

\bibitem{thorne}
  K.~S.~Thorne,
  Astrophys.\ J.\  {\bf 191}, 507 (1974).

\bibitem{sh05}
  S.~L.~Shapiro,
  Astrophys.\ J.\  {\bf 620}, 59 (2005).

\bibitem{marta}
  M.~Volonteri, P.~Madau, E.~Quataert and M.~J.~Rees,
  Astrophys.\ J.\  {\bf 620}, 69 (2005).

\bibitem{io}
  C.~Bambi and K.~Freese,
  Phys.\ Rev.\  D {\bf 79}, 043002 (2009);
  C.~Bambi, K.~Freese, T.~Harada, R.~Takahashi and N.~Yoshida,
  Phys.\ Rev.\  D {\bf 80}, 104023 (2009);
  C.~Bambi, T.~Harada, R.~Takahashi and N.~Yoshida,
  Phys.\ Rev.\  D {\bf 81}, 104004 (2010);
  C.~Bambi and N.~Yoshida,
  Phys.\ Rev.\  D {\bf 82}, 064002 (2010);
  C.~Bambi and N.~Yoshida,
  Phys.\ Rev.\  D {\bf 82}, 124037 (2010).

\bi{mn}
  V.~S.~Manko and I.~D.~Novikov,
  Class.\ Quant.\ Grav.\  {\bf 9}, 2477 (1992).

\bibitem{baba}
  C.~Bambi and E.~Barausse,
  Astrophys.\ J.\  {\bf 731}, 121 (2011).

\bibitem{neutron}
  W.~G.~Laarakkers and E.~Poisson,
  Astrophys.\ J.\  {\bf 512}, 282 (1999).

\bibitem{bp1}
  P.~A.~G.~Scheuer and R.~Feiler,
  Mon.\ Not.\ Roy.\ Astron.\ Soc.\  {\bf 282}, 291 (1996).

\bibitem{bp2}
  P.~Natarajan and J.~E.~Pringle,
  Astrophys.\ J.\  {\bf 506}, L97 (1998).

\bibitem{barausse}
  P.~Pani, E.~Barausse, E.~Berti and V.~Cardoso,
  Phys.\ Rev.\  D {\bf 82}, 044009 (2010).

\bibitem{kokkotas}
  N.~Andersson, D.~I.~Jones, K.~D.~Kokkotas and N.~Stergioulas,
  Astrophys.\ J.\  {\bf 534}, L75 (2000).

\bibitem{bumpy}
  N.~A.~Collins and S.~A.~Hughes,
  Phys.\ Rev.\  D {\bf 69}, 124022 (2004);
  K.~Glampedakis and S.~Babak,
  Class.\ Quant.\ Grav.\  {\bf 23}, 4167 (2006);
  J.~R.~Gair, C.~Li and I.~Mandel,
  Phys.\ Rev.\  D {\bf 77}, 024035 (2008);
  T.~A.~Apostolatos, G.~Lukes-Gerakopoulos and G.~Contopoulos,
  Phys.\ Rev.\ Lett.\  {\bf 103}, 111101 (2009);
  S.~J.~Vigeland and S.~A.~Hughes,
  Phys.\ Rev.\  D {\bf 81}, 024030 (2010).

\bibitem{vsop}
  C.~Bambi and N.~Yoshida,
  Class.\ Quant.\ Grav.\  {\bf 27}, 205006 (2010);
  T.~Johannsen and D.~Psaltis,
  Astrophys.\ J.\  {\bf 718}, 446 (2010).

\bibitem{barack}
  L.~Barack and C.~Cutler,
  Phys.\ Rev.\  D {\bf 75}, 042003 (2007).

\bibitem{mm}
  M.~Calvani, R.~Catenacci and F.~Salmistraro,
  Lett.\ Nuovo Cim.\  {\bf 16}, 460 (1976);
  V.~S.~Manko and C.~Moreno,
  Mod.\ Phys.\ Lett.\  A {\bf 12}, 613 (1997).

\end{thebibliography}
\end{document}